# PENERAPAN KNOWLEDGE MANAGEMENT SYSTEM SALES AND CUSTOMER CARE PADA PT SATRIA MEDIKANTARA PALEMBANG

**Yandi Pranata[1], Leon A. Abdillah[2], Siti Sa'uda[3]**
Fakultas Ilmu Komputer, Universitas Bina Darma
Email: yandipranata@gmail.com[1], leon.abdillah@yahoo.com[2*], siti_sauda@binadarma.ac.id[3]

**ABSTRAK**

*PT Satria Medikantara Palembang has a great desire to apply knowledge management system, therefore the documentation of knowledge and its utilization needs to be well managed in the context of performance improvement. The implementation of knowledge management in PT Satria Medikantara Palembang is considered very good and can have a positive impact for the quality of employees. Where every employee can store and document and share knowledge owned, so that other employees can access, even learn and discuss with other employees based on the knowledge posted. Then when needed a knowledge, it is very easy to find in the database searching feature in the web based knowledge management system at PT Satria Medikantara Palembang. The methodology used in this study refers to the methodology of knowledge management system life cycle developed by Awad and Ghaziri (2010), and this system will be based on web using PHP programming language.*

**Kata kunci:** Customer Care, Knowledge Managing Systems, Sales, PT Satria Medikantara.

## 1. PENDAHULUAN

Kondisi persaigan usaha dan perkembangan teknologi infomrasi (TI) dan sistem informasi (SI) telah memicu perusahaan-perusahaan untuk mengambil langkah-langkah yang tepat dan inovatif. Institusi maupun lembaga memerlukan suatu inovasi baru dalam menyusun strategi untuk menjaga kelangsungan bisnis dan dapat bertahan dari segala ancaman yang ada. Salah satu solusi yang dapat digunakan untuk bertahan dan bersaing adalah dengan mengelola pengetahuan yang ada. Instansi yang bertahan lama adalah instansi yang memiliki kemampuan untuk mengelola pengetahuan (*knowledge*) perusahaannya dengan menggunakan teknologi terkini [1].

*Knowledge Management* (KM) dapat digambarkan sebagai praktik menangkap *tacit knowledge* dan mengubahnya menjadi *explicit knowledge* [2]. Dalam suatu perusahaan atau instansi, *information and knowledge* (IK) disimpan di setiap individu dalam organisasi dalam bentuk pengalaman, keterampilan, dll [3]. Sedangkan KMS adalah alat yang ditujukan untuk mendukung *knowledge management* [4]. Pengetahuan berkembang sejalan dengan kebutuhan dan tantangan untuk masa depan yang lebih baik [5].

Ada sejumlah penelitian terdahulu yang penulis jadikan rujukan pada pengembangan KMS ini. Penelitian-peneliitan tersebut, yaitu: 1) Rancang bangun KMS di SMA [6] untuk pendokumentasiana knowledge yang dimiliki oleh pegawai, 2) Penerapan KMS *Sales and Customer Care* [7] untuk menyimpan *knowledge* dengan tampilan yang lebih *friendly*, 2) Implementasi KMS dalam Rekonsilasi Aset [8] dilakukan untuk mengatasi masalah kesulitan dalam hal mengumpulkan data dalam bentuk informasi karena data ditempatkan di *folder* yang berbeda, sehingga pengetahuan tentang aset sulit ditemukan, dan 3) Evaluasi Infrastruktur *Knowledge Sharing* Pegawai [9] untuk menanggulangi sering terjadinya pengulangan kesalahan-kesalahan yang pernah dilakukan sebelumnya serta untuk memfasilitasi masalah pendokumentasian serta meningkatkan kualitas kerja pegawai.

PT Satria Medikantara Palembang merupakan perusahaan swasta di Kota Palembang yang bergerak di bidang penjualan alat-alat keseahtan dan obat-obatan. Perusahaan ini memiliki keinginan yang besar untuk turut menerapkan *knowledge management* di bagian *Sales and Customer Care*. Oleh karena itu pendokumentasian pengetahuan dan pemanfaatannya perlu dikelola dengan baik dalam konteks peningkatan kinerja. Diterapkannya *Knowledge Management System* (KMS) di bagian *Sales and Customer Care* ini dinilai sangat baik dan dapat berdampak positif bagi kualitas pegawai. Pegawai di bagian *Sales and Customer Care* dapat menyimpan dan mendokumentasikan serta *sharing* pengetahuan yang dimiliki, sehingga pegawai pada bagian *Sales and Customer Care* dapat mengakses, mempelajari dan dapat berdiskusi antara sesama pegawai berdasarkan pengetahuan yang di-*posting*.

Berdasarkan wawancara pada pimpinan, bagian *Sales and Customer Care* pada PT Satria Medikantara Palembang, selama ini pengolahan *knowledge* dalam perusahaan ini masih tidak terdistribusi dengan baik. Khususnya untuk bagian *Sales and Customer Care* kesulitan dalam hal penyimpanan *knowledge* karena tidak adanya sebuah sistem yang dapat menyimpan *knowledge* dengan tampilan KMS yang lebih *friendly*. Terjadi perulangan pembahasan masalah dan solusi yang sama pada saat *meeting*, sering kehilangan data, kesulitan dalam pendokumentasian data pengetahuan, *sharing knowledge* tidak efisien. Hal inilah yang menyebabkan penerapan KMS menjadi penting di bagian Sales and Customer PT Satria Medikantara Palembang. Dari uraian diatas penulis mencoba untuk merancang sistem yang dapat membantu pengelolaan pengetahuan pada Bagian *Sales and Customer Care* PT Satria Medikantara Palembang.

## 2. METODOLOGI PENELITIAN

### 2.1 Objek Penelitian

Objek pada penelitian ini adalah PT Satria Medikantara yang berlokasi di Jalan Mayor Ruslan No. 30 B Palembang. PT ini didirikan pada tanggal 24 Juli 1991 berdasarkan akte notaris Robert Tjahjaindra, S.H. Nomor 120-1991. Perusahaan ini merupakan perusahaan yang bergerak di bidang usaha penjualan obat-obatan dan alat-alat kesehatan khususnya sebagai pedagang besar farmasi (PBF). Permodalan yang digunakan dalam melaksanakan operasi usahanya adalah modal sendiri dan sebagian lagi dengan pembiayaan secara kredit. Bangunan PT Satria Medikantara yang berlantai dua ini mempunyai luas tanah dan bangungan sebesar 325 m2 dan 280 m2. PT Satria Medikantara didirikan oleh Bapak Drs. Sathony Kaslim.

Pada tahap awal operasinya PT Satria Medikantara hanya mampu melayani permintaan dari dalam kota Palembang saja. Tetapi seiring perkembangan selanjutnya sudah mampu untuk melayani permintaan dari luar kota Palembang.

### 2.2 Metode Pengumpulan Data

Pada penelitian ini, tim peneliti melibatkan 2 (dua) metode pengumpulan data, yaitu data primer dan data sekunder. Data primer adalah secara langsung diambil dari objek penelitian oleh peneliti baik perorangan maupun organisasi, terdiri dari: 1) Wawancara. Metode ini melakukan pengumpulan data melalui tanya jawab langsung dengan pihak pihak yang berkaitan erat dengan permasalahan yang diteliti. Dalam penelitian ini, wawancara dilakukan dengan pimpinan, bagian *sales and customer care* pada PT Satria Medikantara Palembang, dan 2) Observasi ke perusahaan, terutama bagian sales and customer care. Data sekunder adalah data yang didapat secara tidak langsung dari objek penelitian. Peneliti mendapatkan data yang sudah jadi yang dikumpulkan oleh pihak lain dengan berbagai cara atau metode baik secara komersial maupun non komersial, terdiri dari: 1) Studi pustaka dengan melakukan pencarian bahan pendukung seperti buku, internet, dan media informasi yang berhubungan dengan objek penelitian, dan 2) Laporan yang ada pada PT Satria Medikantara Palembang.

### 2.3 Metode Pengembangan Sistem

Dalam penelitian ini metode pengembangan *knowledge management systems* yang digunakan adalah adopsi metode *Knowledge Management System Life Cycle* (KMSLC) [10] yang memilik 6 (enam) tahapan berikut: 1) *Evaluate Existing Infrastructure*, 2) *Form the KM Team*, 3) *Knowledge Capture*, 4) *Design KM Blueprin*t, 5) *Verify and Validate the KM System*, dan 6) *Implement the KM System*.

Rancangan *use case* atau diagram *use case* merupakan pemodelan untuk kelakuan (*behavior*) sistem informasi yang akandibuat. *Use case* mendeskripsikan sebuah interaksi antara satu atau lebih aktor dengan sistem informasi yang akan dibuat. Secara umum *use case* digunakan untuk mengetahui fungsi apa saja yang ada di dalam sebuah sistem informasi dan siapa saja yang berhak menggunakan fungsi-fungsi itu. Gambar 1 menampilkan diagram *use case* KMS pada penelitian ini.

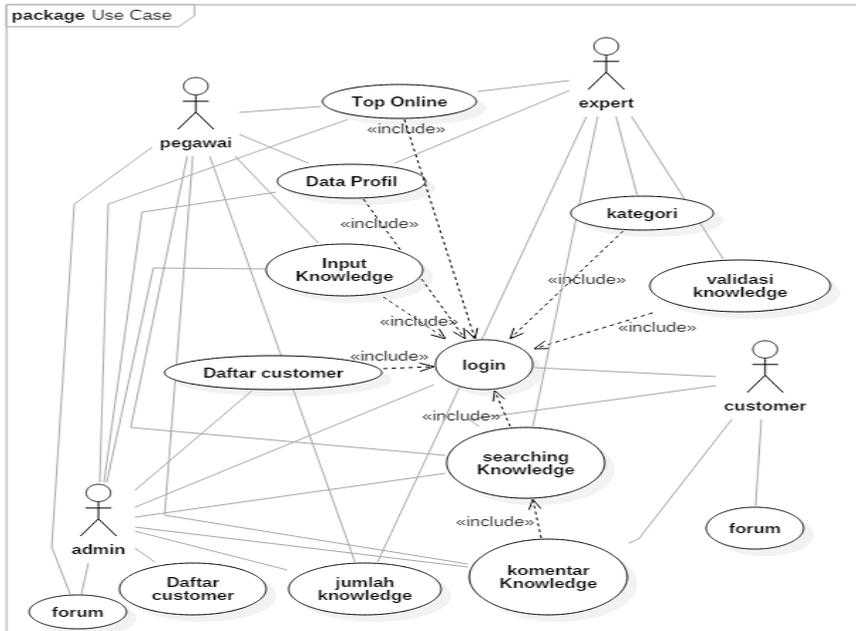

**Gambar 1.** *Use Case Diagram*

## 3. HASIL DAN PEMBAHASAN

Hasil terakhir yang didapat pada sistem ini adalah *Knowledge Management System Sales and Customer Care* pada PT Satria Medikantara Palembang. Sistem ini diperuntukkan 4 (empat) jenis *user*, yaitu *expert*, *admin*, pegawai dan *customer*. Sistem ini memiiki beberapa file-file, yang berupa halaman-halaman antar muka yang masing-masing memiliki menu yang disesuaikan dengan rancangan sistem yang telah dibahas sebelumnya.

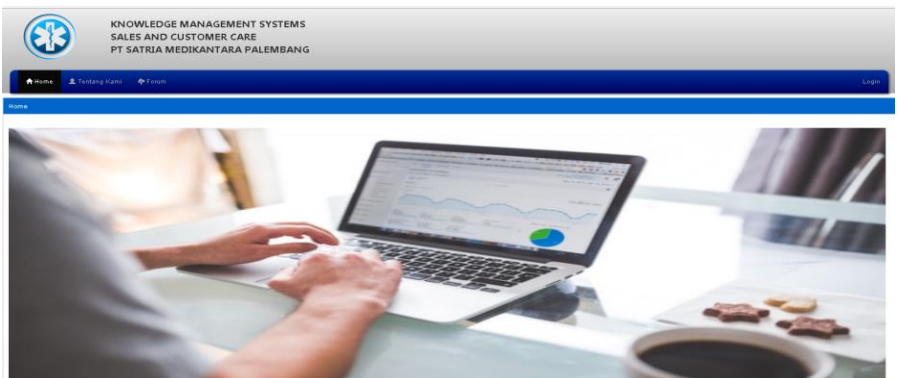

**Gambar 2.** Menu Utama

### 3.1 Menu Utama

Halaman utama akan tampil pada saat membuka portal web sebelum user melakukan pengolahan data, terdapat menu home, tentang kami (profil perusahaan, visi dan misi dan *contact person*), forum untuk melakukan diskusi dan menu login untuk user mengakses *website* (lihat gambar 2).

### 3.2 *Input Knowledge*

Halaman menu *Input Knowledge* merupakan halaman yang berfungsi untuk menambahkan pengetahuan yang telah di-*input* oleh *user*. Pada menu *input knowledge* berisikan daftar *knowledge* yang telah kita *input* yang terdiri dari nomor, kategori, sifat, judul, isi *knowledge*, status pending/*valid*, menu *action* untuk mengedit atau menghapus dan menu tambah data yang berguna untuk menambahkan *knowledge* yang baru (gambar 3).

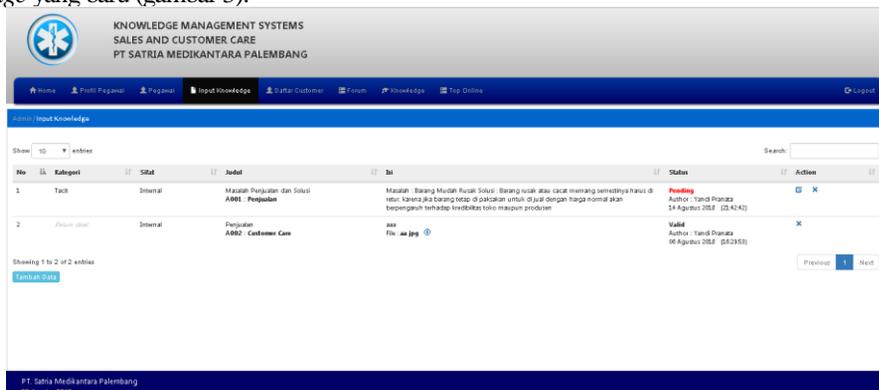

**Gambar 3.** Menu *Input Knowledge*

### 3.3 Daftar *Knowledge*

Halaman daftar *knowledge* merupakan halaman yang berisikan daftar-daftar pengetahuan yang telah di-*input* oleh *user*. Di menu daftar *knowledge* ini terdapat menu *searching knowledge* yang berfungsi untuk mencari pengetahuan yang ingin dicari oleh *user* (gambar 4).

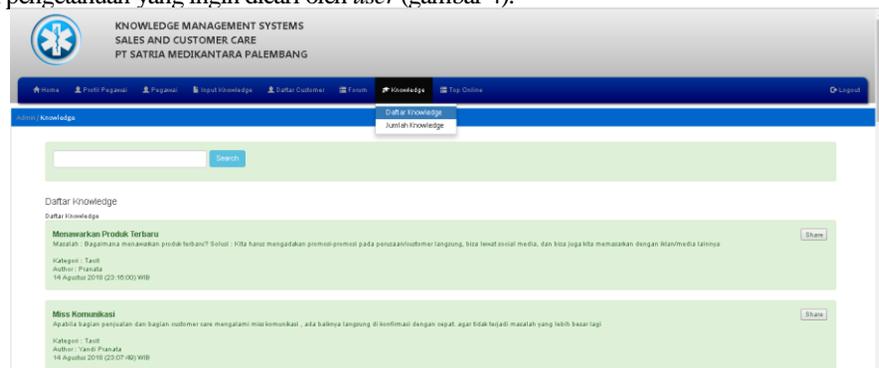

**Gambar 4.** Menu *Daftar Knowledge*

### 3.4 Menu Jumlah *Knowledge*

Halaman menu jumlah *knowledge* merupakan halaman yang berisikan jumlah keseluruhan *knowledge* yang ada. Jumlah *knowledge* yang ada digambarkan dalam diagram batang dan dilengkapi dengan jumlah *knowledge* dibagian atas diagram batang-nya.

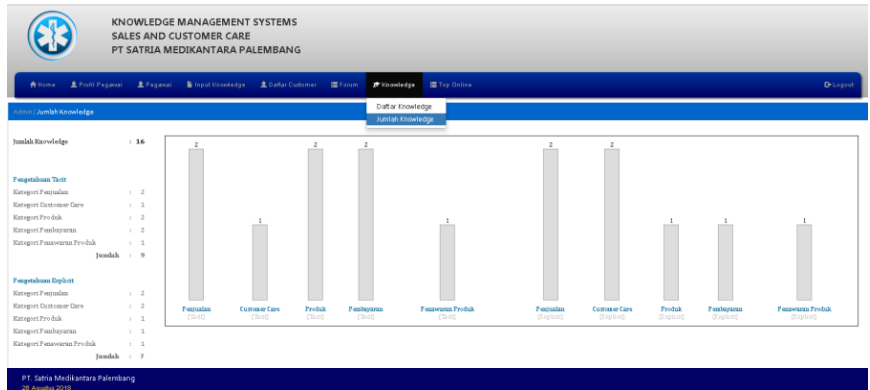

**Gambar 3.** Menu *Daftar Knowledge*

## 4. KESIMPULAN DAN SARAN

Berdasarkan pembahasan dan uraian pada bab-bab sebelumnya, maka penulis menarik kesimpulan sebagai berikut: 1) KMS yang telah dibangun pada PT Satria Medikantara Palembang dapat membantu karyawan dalam proses *discovery knowledge* yaitu mendokumentasikan data pengetahuan agar dapat disimpan dan dibagi dengan pegawai lainnya, 2) Dengan adanya fitur pencarian (*search knowledge*), memudahkan pegawai dalam mencari pengetahuan yang mereka inginkan sesuai dengan *keyword* yang di-*input*-kan, dan 3) KMS ini mampu dijadikan sarana berdiskusi pengetahuan atas permasalahan yang ada antar pegawai *sales and customer care* dengan memanfaatkan kolom komentar pada pengetahuan yang ada.

Pembahasan KMS pada penelitian ini masih terbatas pada bagian *sales and customer care*, diharapkan kedepannya dapat dikembangkan untuk bagian-bagian lainnya di PT Satria Medikantara Palembang. Selain itu, sistem yang dibangun juga diharapkan bisa untuk aplikasi berbasis *smartphone*.

## DAFTAR PUSTAKA